\documentclass[conference]{IEEEtran}
\IEEEoverridecommandlockouts

\usepackage{cite}
\usepackage{amsmath,amssymb,amsfonts}
\usepackage{algorithmic}
\usepackage{hyperref}
\usepackage{graphicx}
\usepackage{textcomp}
\usepackage{xcolor}
\usepackage{gensymb}
\usepackage{comment}
\usepackage{multirow}  
\usepackage{booktabs}   

\def\BibTeX{{\rm B\kern-.05em{\sc i\kern-.025em b}\kern-.08em
    T\kern-.1667em\lower.7ex\hbox{E}\kern-.125emX}}
\begin{document}

\title{TSPC-PFD: TSPC-Based Low-Power High-Resolution CMOS Phase Frequency Detector\\
\thanks{This research was funded in part by National Science Foundation (NSF) award number: 2138253, Rezonent Inc. award number: CORP0061, and UMBC Startup Fund.}
}

\author{\IEEEauthorblockN{Dhandeep Challagundla\textsuperscript{1}, Venkata Krishna Vamsi Sundarapu\textsuperscript{1}, Ignatius Bezzam\textsuperscript{2}, Riadul Islam\textsuperscript{1}}
\IEEEauthorblockA{\textit{\textsuperscript{1}University of Maryland, Baltimore County, \textsuperscript{2}Rezonent Inc.} \\
vd58139@umbc.edu, vsundar1@umbc.edu, i@rezonent.us, riaduli@umbc.edu}

}

\maketitle

\begin{abstract}
Phase Frequency Detectors (PFDs) are essential components in Phase-Locked Loop (PLL) and Delay-Locked Loop (DLL) systems, responsible for comparing phase and frequency differences and generating up/down signals to regulate charge pumps and/or, consequently, Voltage-Controlled Oscillators (VCOs). Conventional PFD designs often suffer from significant dead zones and blind zones, which degrade phase detection accuracy and increase jitter in high-speed applications. This paper addresses PFD design challenges and presents a novel low-power True Single-Phase Clock (TSPC)-based PFD. The proposed design eliminates the blind zone entirely while achieving a minimal dead zone of 40 ps. The proposed PFD, implemented using TSMC 28 nm technology, demonstrates a low-power consumption of $4.41\mu W$ at 3 GHz input frequency with a layout area of $10.42\mu m^2$.  
\end{abstract}


\section{Introduction}

Phase-Locked Loops (PLLs) and Delay-Locked Loops (DLLs) are fundamental clock synchronization circuits widely used in high-speed digital and communication systems. As shown in Figure~\ref{fig:dll_pll}, a DLL (Figure~\ref{fig:dll_pll}(a)) employs a voltage-controlled delay line to align the output phase with the reference clock, ensuring precise delay matching without frequency synthesis. In contrast, a PLL (Figure~\ref{fig:dll_pll}(b)) utilizes a Voltage-Controlled Oscillator (VCO) to generate a frequency-locked output that tracks the reference clock, enabling frequency multiplication and jitter minimization~\cite{razavi2003simple, islam2018hcdn}. 

A crucial component in both PLL and DLL architectures is the Phase Frequency Detector (PFD). The PFD compares the phase and frequency difference between the reference and feedback signals, generating ``Up'' and ``Down'' control signals that regulate the charge pump and loop filter, ultimately adjusting the delay line in DLLs or the VCO in PLLs. The accuracy and speed of the PFD directly impact the loop's lock time, jitter performance, and overall system stability, making its design optimization critical.

Different types of PFDs are designed to meet specific performance requirements. Latch-based PFDs~\cite{koithyar2018faster} are preferred for low-power applications but suffer from dead zone limitations. In contrast, tri-state PFDs help mitigate dead zone effects and have a linear phase-detecting range~\cite{hsu2014sub,crawford1994frequency}.

PFD design improvements primarily target reducing the dead zone and minimizing phase noise, while balancing power consumption and operating frequency~\cite{Huo_pfd:2024}. Unlike conventional synchronous systems, where Power, Performance, and Area (PPA) optimization is standard~\cite{Islam_cmcs:2017, Jenila_ppa:2025, Islam_asicon:2011, Kundu_ppa:2025, guthaus2017current, Islam_ncfet:2021}, achieving effective PPA optimization for a PFD remains non-trivial. Various techniques, such as reset-path removal and the use of True Single-Phase Clock (TSPC)~\cite{rabaey2002digital, islam2011high, challagundla2023design,islam2024system} logic, offer promising solutions, but each comes with trade-offs. The selection of an appropriate PFD architecture depends on the specific application requirements, particularly in high-performance PLLs where jitter and phase accuracy are critical considerations.

This work proposes a novel TSPC-based PFD circuit designed to enhance power efficiency and reduce dead-zone effects, thereby improving the locking characteristics of both PLL and DLL architectures in advanced low-power applications. The key contributions of this research are:

\begin{figure}[t!]
\centerline{\includegraphics[width = 0.4\textwidth]{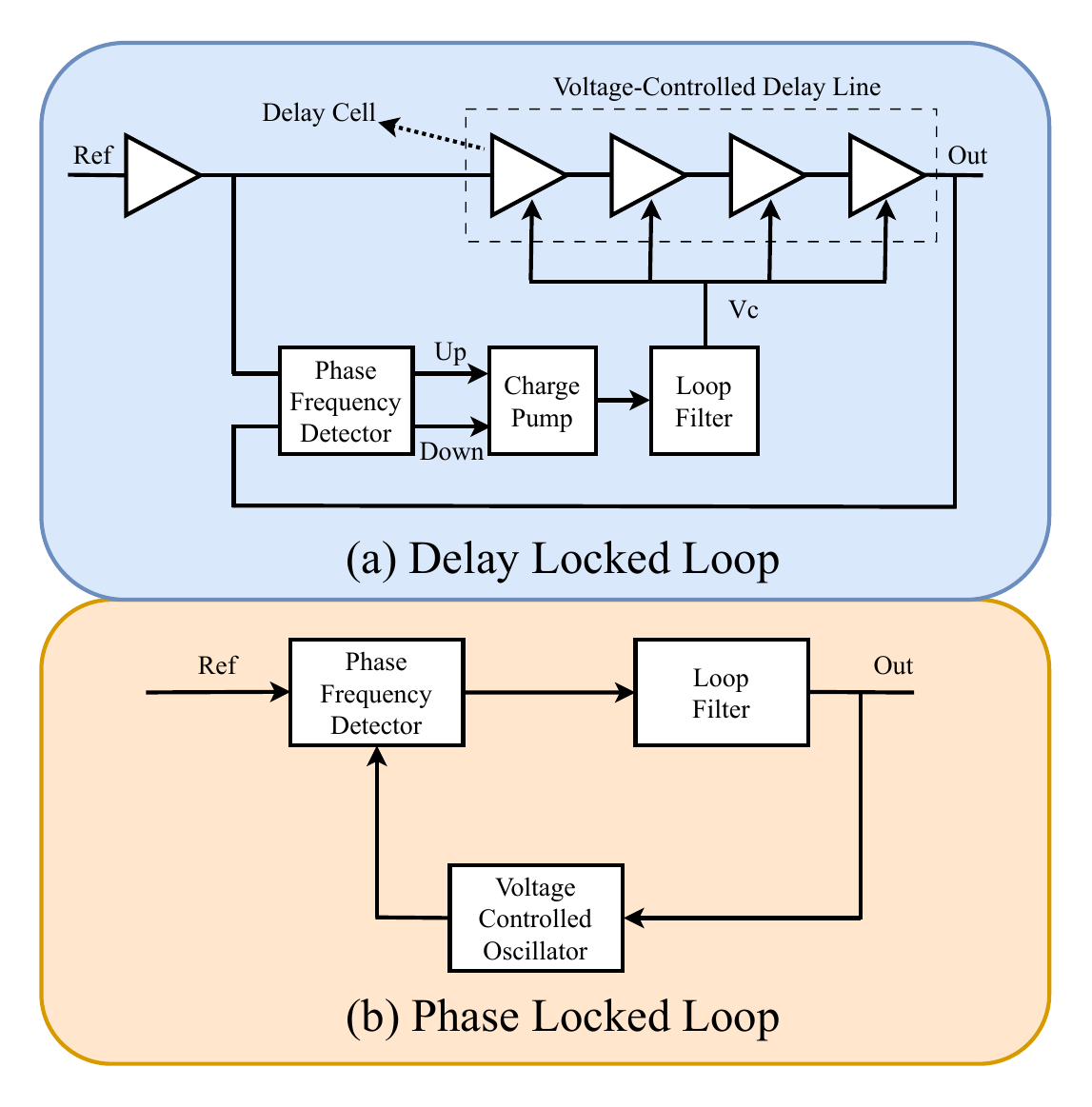}}
\vspace{-0.27cm}
\caption {Generalized block diagram of (a) a delay-locked loop and (b) a phase-locked loop~\cite{rabaey2002digital}.}
\label{fig:dll_pll}
\vspace{-0.5cm}
\end{figure}
 
\begin{itemize}
    \item A novel low-power, high-resolution PFD circuit based on TSPC logic.
    \item The dead zone of the proposed PFD is primarily determined by the setup time of the TSPC latch, exhibiting a minimal dead zone of 40 ps.
    \item The proposed design achieves $2.5\times$ lower power consumption compared to \cite{divya:2024}, and a 72.5\% reduction in power relative to \cite{rezaeian:2020}.
\end{itemize}
\section{Background}

Traditional PFDs suffer from a large dead zone in their phase detection characteristics at a steady state, leading to increased jitter in the locked state~\cite{sharma2021brief}. This dead zone occurs when the phase difference between the input signals is too small for the PFD to detect, causing erroneous phase detection and potentially locking the PLL to an incorrect phase. Additionally, conventional tristate charge pumps used in PLLs introduce issues such as charge sharing, charge injection, and clock feedthrough when the $Up$ and $Down$ signals switch states, further degrading the performance by introducing phase noise.

Another limitation of conventional PFD designs is the blind zone, which occurs when the input signals are close to ±360°. This can lead to instability in phase detection, further complicating the lock-in process of the PLL.

\begin{figure}[t]
\centerline{\includegraphics[width = 0.4\textwidth]{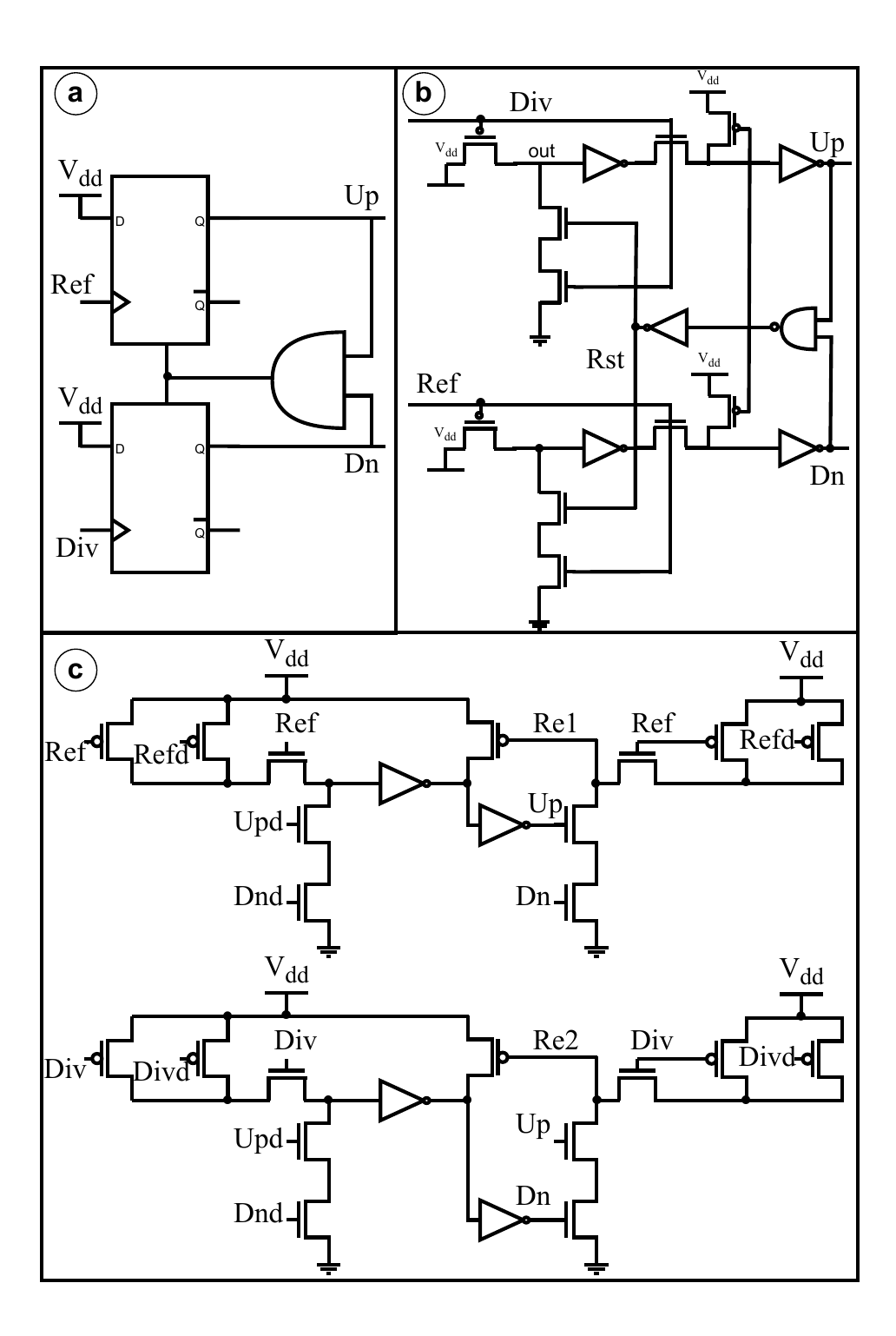}}
\vspace{-0.27cm}
\caption {(a) Traditional PFD using D latch design, (b)a pass transistor based fast PFD~\cite{mansuri:2001} (c) existing dead zone free PFD~\cite{divya:2024} using 22 transistors.}
\label{fig:comparisons}
\vspace{-0.05cm}
\end{figure}

Various design modifications have been explored to minimize the dead zone~\cite{azadmousavi2018novel,divya:2024,Sharma:2024,rezaeian:2020, ismail2009cmos, liang202240, park2021design, divya2023dead, ju202148gb} while maintaining power efficiency~\cite{jamadi_24_sram, abdul2017low} and high-speed operation~\cite{badiger2024design,lee1999high, pradhan2021design}. One common approach involves using D flip-flops and an ``AND" gate to generate the $Up$ and $Down$ signals~\cite{rabaey2002digital, nagarajan2022delay}. While this method is simple, it suffers from high power consumption at increased operating frequencies and requires a large transistor count, leading to higher area utilization.

Figure~\ref{fig:comparisons} illustrates various PFD architectures. Figure~\ref{fig:comparisons}(a) shows the conventional PFD design employing D--latches and a reset logic gate, which suffers from a considerable dead zone due to latch timing limitations. Figure~\ref{fig:comparisons}(b)~\cite{mansuri:2001} presents a PFD optimized for low signal-to-noise ratio (SNR) environments, offering improved sensitivity but with limited dead zone mitigation. Figure~\ref{fig:comparisons}(c) depicts an advanced dead zone-free PFD architecture, as proposed in \cite{divya:2024}, utilizing 22 transistors and incorporating inverters for faster reset paths to minimize the dead zone.

An alternative design aims to eliminate the dead zone entirely~\cite{Sharma:2024} by removing the reset path in the PFD. This results in a more linear phase detection characteristic, enabling precise phase error correction. Such an approach has demonstrated a maximum operating frequency of 4 GHz with high power consumption.

Another notable design utilizes a D flip-flop combined with gate diffusion input (GDI) logic technique to optimize power efficiency~\cite{rezaeian:2020}. This PFD offers significantly lower power consumption while achieving operating frequencies of up to 3.33 GHz. However, it suffers from a high blind zone of $2\pi/10$ around $\pm360\degree$, which impacts performance under certain operating conditions. Moreover, previous works have demonstrated their designs at higher technology nodes. In contrast, this work presents a novel TSPC-based PFD circuit implemented in 28nm technology, achieving zero blind zone, minimal dead zone, and low power consumption. The TSPC latch is advantageous as it operates with a single clock phase, eliminating the need for complementary clocks. Its dynamic structure enables faster switching and a lower transistor count, making it well-suited for high-frequency, low-power PFDs.

\section{Proposed Phase Frequency Detector Circuit}
\begin{figure}[t]
\centerline{\includegraphics[width = 0.5\textwidth]{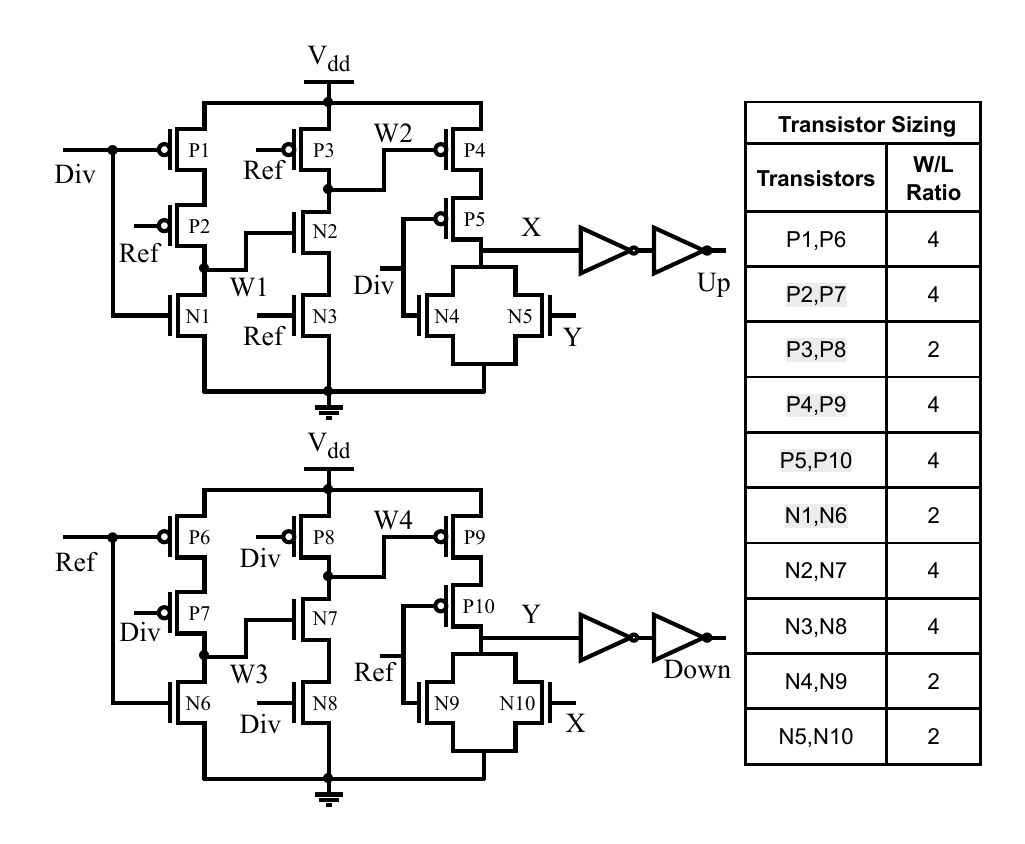}}
\vspace{-0.27cm}
\caption {The proposed TSPC-based phase frequency detector circuit is designed using 20 transistors with symmetric design to generate $Up$ and $Down$ signals.}
\label{fig:schematic}
\vspace{-0.05cm}
\end{figure}
The PFD is a crucial component in PLLs and DLLs and is responsible for detecting both phase and frequency differences between the reference clock (``Ref'') and the feedback clock (``Div''). The proposed PFD, shown in Figure~\ref{fig:schematic}, is implemented using a CMOS-based dynamic logic design to achieve high-speed operation with minimal power consumption. The circuit consists of two symmetrical dynamic TSPC flip-flop~\cite{islam2018low}-like structures, each dedicated to generating $Up$ and $Down$ signals based on the relative timing of ``Ref'' and ``Div.'' These signals serve as inputs to a charge pump, ultimately controlling the loop filter and frequency synthesis mechanism.

The PFD operates by comparing the phase and frequency of ``Ref'' and ``Div'' and producing an appropriate control signal. When the reference clock leads the divided clock, the $Up$ signal is asserted, prompting the charge pump to increase the control voltage of the VCO or delay line. Conversely, when the feedback clock leads the reference clock, the $Down$ signal is activated, instructing the system to slow down. The circuit ensures that $Up$ and $Down$ signals are never active simultaneously, preventing ambiguous states and reducing phase jitter. Additionally, an internal reset mechanism is implemented to suppress excessive pulses, thereby improving the overall locking stability of the loop.

\begin{figure}[t]
\centerline{\includegraphics[width = 0.5\textwidth]{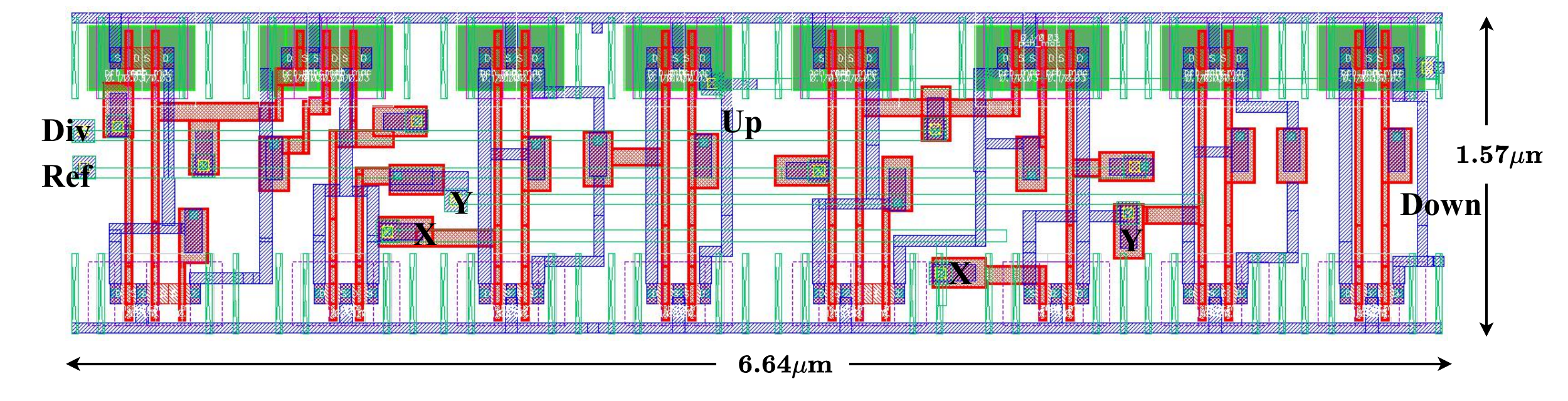}}
\vspace{-0.27cm}
\caption {The layout of the proposed TSPC-based phase frequency detector circuit occupies $10.42\, \mu m^2$ and uses metal 1 to metal 3 layers.}
\label{fig:layout}
\vspace{-0.5cm}
\end{figure}

\begin{figure}[t]
\centerline{\includegraphics[width = 0.45\textwidth]{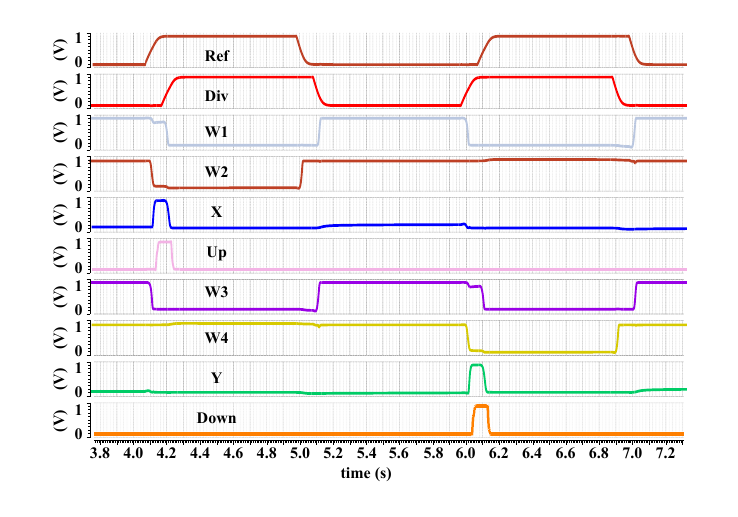}}
\vspace{-0.27cm}
\caption {Simulation results of the proposed PFD circuit producing appropriate up and down signals according to the phase difference of ``Ref'' and ``Div'' signals.}
\label{fig:simulation}
\vspace{-0.5cm}
\end{figure}

The circuit schematic of the proposed PFD is shown in Figure~\ref{fig:schematic}, and the corresponding simulation waveform is described in Figure~\ref{fig:simulation}. When both ``Div'' and ``Ref'' signals are ``0,'' transistors P1 and P2 are ``ON,'' which charges the wire W1 to $V_{dd}$ and, in turn, activates N2. As ``Ref'' transitions to ``1'' while ``Div'' remains ``0,'' node X is driven to ``1.'' Since N2 remains ``ON'' and N3 also turns ``ON'' (``Ref'' = ``1''), wire W2, the input to P4 discharges. Simultaneously, as ``Div'' remains ``0,'' P5 stays ``ON,'' allowing X to charge and maintain its state at ``1.'' With X = ``1,'' transistor N10 becomes ``ON,'' discharging node Y and setting Y to ``0.'' This is shown in the first cycle of the simulation waveform, Figure~\ref{fig:simulation}. 

In the second cycle, as shown in Figure~\ref{fig:simulation}, when ``Div'' transitions to ``1'' while ``Ref'' remains ``0,'' transistors N8 and N7 turn ``ON,'' discharging the gate input of P9. This causes Y to be set to ``1,'' which subsequently turns N5 ``ON,'' driving X to ``0.'' This operational sequence ensures that the circuit accurately detects the phase difference between the ``Ref'' and ``Div'' signals, enabling reliable phase frequency detection essential for high-speed clock synchronization applications.

Figure~\ref{fig:layout} shows the layout of the proposed PFD circuit. It has a standard cell height of $1.57 \mu m$ and an area of $10.42\mu m^2$ and is constructed using Metal 1 to Metal 3 layers.

\section{ Results and Discussion}

\subsection{Experimental Setup}

The proposed TSPC-based PFD circuit was designed using TSMC 28nm CMOS technology. The schematic and layout designs were implemented in Cadence Virtuoso, and transient simulations were conducted using the Cadence Spectre simulator. Simulations were performed with a 1 GHz to 3 GHz clock frequency, and power consumption values were extracted directly from the Spectre tool. Since 28 nm implementations of conventional PFDs were not readily available, we used Dennard's power scaling law~\cite{Dennard:1974} to compare the reference circuits. Additionally, transfer characteristic simulations were performed to determine the dead zone of the proposed circuit. To assess design robustness, process, voltage, and temperature (PVT) variation analyses were also conducted.

\subsection{Transfer Characteristics}

\begin{figure}[t]
\centerline{\includegraphics[width = 0.4\textwidth]{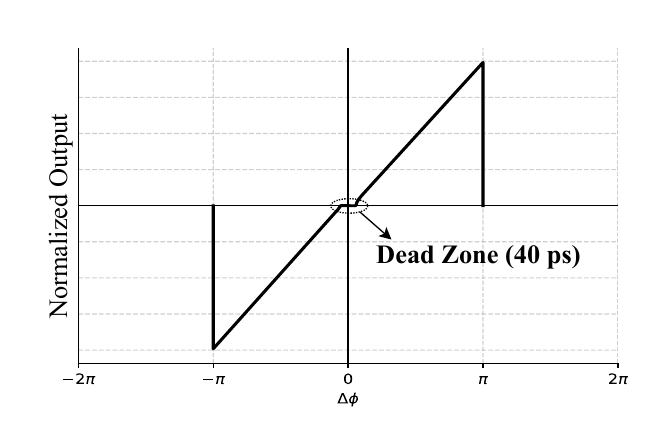}}
\vspace{-0.27cm}
\caption {The transfer characteristics of the proposed TSPC-based phase frequency detector circuit demonstrate a low dead zone of 40 ps.}
\label{fig:transfer}
\vspace{-0.5cm}
\end{figure}

The transfer characteristics plot shown in Figure~\ref{fig:transfer} is obtained by simulating the \textit{Ref} and \textit{Div} signals with a phase difference ranging from $-\pi$ to $\pi$. The plot illustrates a normalized output response, demonstrating the expected phase detection behavior. Notably, the frequency divider operation observes a dead zone of 40 ps. This dead zone is relatively low due to the dependence of the PFD on the TSPC latch, where the setup time of the flip-flop introduces this phase-detection limitation. The minimized dead zone ensures improved phase tracking accuracy and enhances the overall performance of the PLL system. 

\subsection{Comparison with Previous Works}

\begin{table}[h]

\centering
    \caption{Comparison of various PFD circuits}
    \renewcommand{\arraystretch}{1.5} 
    \resizebox{\columnwidth}{!}{%
    \fontsize{14pt}{16pt}\selectfont
\begin{tabular}{|l|c|c|c|c|c|c|c|}
\hline
 & \textbf{This Work} & \cite{Sharma:2024} & \cite{divya:2024} & \cite{rezaeian:2020} & \cite{mansuri:2001} & \cite{fathi:2019} & \cite{hsu:2014}\\
\hline
\textbf{Process Technology (nm)} & \textbf{28} & 180 & 28 & 28 & 28 & 28 & 65  \\
\hline
\textbf{Dead Zone (ps)} & \textbf{40} & 0 & 0 & 0 & 70 & 45 & -- \\
\hline
\textbf{Power Consumption (uW)} & \textbf{4.41} & 201.6* & 11.04** & 15.44** & 2.92** & 21.9** & 422.4*  \\
\hline
\textbf{Transistor Count} & \textbf{20} & 8 & 22 & 16 & 54 & 22 & 14  \\
\hline
\textbf{Max Frequency (GHz)} & \textbf{3} & 4 & 2.9 & 3.3 & 0.8 & 1 & 2.2  \\ 
\hline
\end{tabular}%
}
\label{tab:comparison}
\vspace{2mm}
\begin{minipage}{\columnwidth}
\centering
\textsuperscript{*} Scaled using Dennard's scaling law~\cite{Dennard:1974} to match the 28nm technology.\\
\textsuperscript{**} Simulated circuits in-house using 28nm technology.
\end{minipage}
\vspace{-0.6cm}
\end{table}

Table~\ref{tab:comparison} presents a comparative analysis of various PFD designs in terms of process technology, dead zone, power consumption, transistor count, and maximum operating frequency. The proposed TSPC-based PFD, implemented in TSMC 28 nm technology, demonstrates a low dead zone of 40 ps while consuming only $4.41\mu W$ of power at 3 GHz. We performed simulations of comparative circuits under identical conditions as those used for the proposed circuit to measure their power consumption. To ensure a fair comparison for remaining circuits, we applied Dennard's power scaling law~\cite{Dennard:1974} to scale down the power consumption results from larger technology nodes to the 28 nm technology node.

Compared to \cite{Sharma:2024}, the proposed design achieves 99.3\% reduction in power consumption. Additionally, the proposed PFD design consumes $2.5\times$ lower power than \cite{divya:2024} and $3.5\times$ lower than \cite{rezaeian:2020}. While these designs have zero dead zones, they have high power consumption. The proposed TSPC-based PFD design exhibits a higher operating frequency and a lower dead zone of 40 ps compared to \cite{mansuri:2001} and \cite{fathi:2019}, which have values of 70 ps and 45 ps, respectively.

\subsection{PVT Variation Analysis}

\begin{figure}[t]
\centerline{\includegraphics[width = 0.4\textwidth]{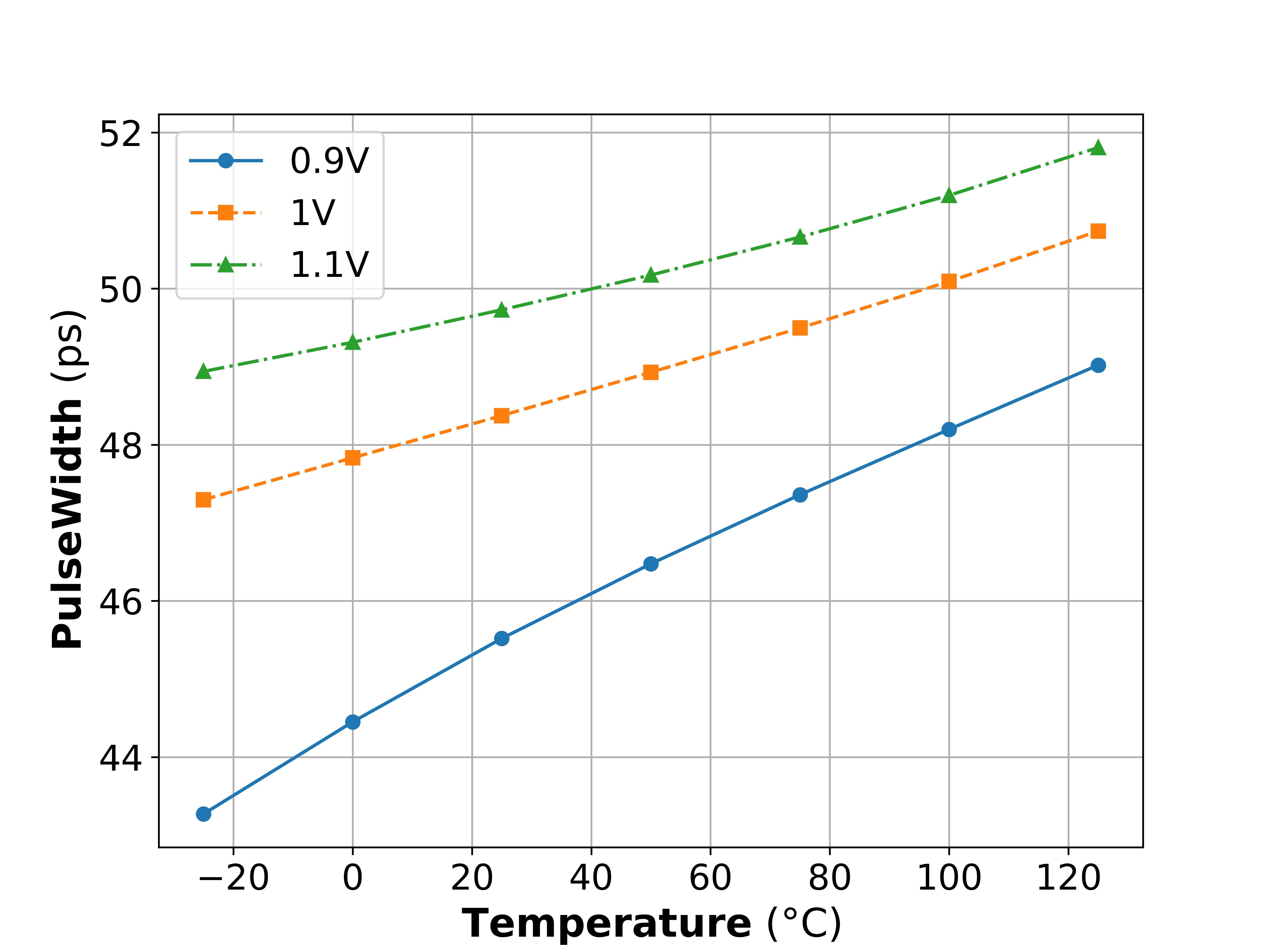}}

\caption {The temperature and voltage variations show minimal changes across temperatures ranging from $-25\degree C$ to $125 \degree C$ and $\pm10\%$ change in supply voltage.}
\label{fig:vt_variations}
\vspace{-0.5cm}
\end{figure}

Figure~\ref{fig:vt_variations} shows the temperature-voltage variation analysis for a nominal phase difference of $\pm0.1\pi$ at 1 GHz, where the ideal pulse width is 50 ps. We considered seven different temperatures ranging from $-25\degree C$ to $125\degree C$ and $\pm10\%$ changes in supply voltages. The average pulse width of ``Up'' and ``Down'' is measured at each combination of temperature and voltage considering a nominal phase difference of $0.1\pi$. It is observed that the proposed PFD circuit generates $1.1\times$ higher pulse width on average with a higher supply voltage of 1.1 V than 0.9 V. Moreover, the proposed circuit generates only 5.5\% higher average pulse width at $100\degree C$ when compared to $25\degree C$. This analysis showcases a maximum change of 44 ps to 52 ps average pulse width when considering extreme temperatures of -$25\degree C$ at 0.9 V and $125\degree C$ at 1.1 V, respectively.

\begin{figure}[t]
\centerline{\includegraphics[width = 0.5\textwidth]{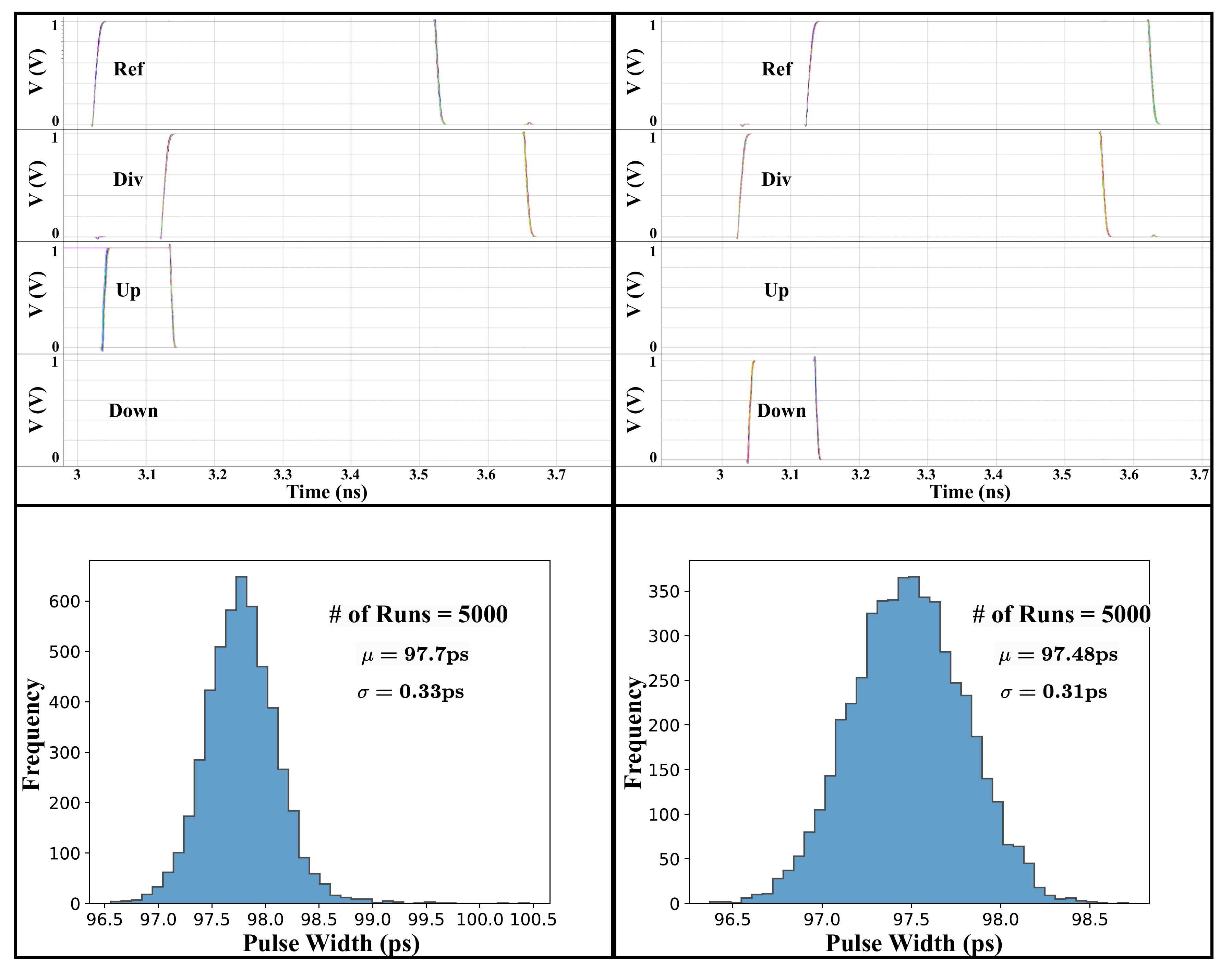}}

\caption {The process variation analysis shows a very similar mean pulse width value when considering 5000 samples of $3\sigma$ deviations in $\pm10\%$ change in length of all transistors of the proposed PFD circuit. }
\label{fig:mc_sim}
\vspace{-0.5cm}
\end{figure}

Figure~\ref{fig:mc_sim} presents the process variation analysis of the proposed PFD circuit, based on 5000 samples of the ``Up'' and ``Down'' signal pulse widths, considering $\pm10\%$ length variation of all transistors under $3\sigma$ deviations. This analysis is performed considering a phase difference of $\pm0.2\pi$ at 1 GHz. The analysis yielded a mean pulse width of 97.7 ps with a standard deviation of 0.33 ps for the ``Up'' signal. Meanwhile, the ``Down'' pulse width exhibits a very similar mean pulse width of 97.48 ps and a standard deviation of only 0.31 ps.
\vspace{-0.2cm}

\section{Conclusion}

In this work, we proposed a novel low-power TSPC-based PFD optimized for high-speed and low-power operation. The proposed design was implemented in TSMC 28 nm CMOS technology and achieved a dead zone of 40 ps with zero blind zones, demonstrating improved phase detection accuracy. Our design achieved a low power consumption of $4.41\mu W$ at 3 GHz frequency, with a layout area of $10.42\mu m^2$. Comparison with prior works shows high power savings ranging from 61\% when compared with \cite{divya:2024} to 80\% when compared with \cite{fathi:2019}. Additionally, PVT variation analysis conducted using 5000 samples of $\pm10\%$ length variation with $3\sigma$ deviations demonstrated the robustness of the proposed PFD circuit with a mean average pulse width of 97 ps at a nominal phase difference of $\pm0.2\pi$ around $0\degree$.

\bibliographystyle{IEEEtran}
\bibliography{main}

\end{document}